# Spin-orbit coupling and weak antilocalization in thermoelectric material $\beta$-K$_2$Bi$_8$Se$_{13}$

J. Hu, J.Y. Liu, and Z.Q. Mao*

Department of Physics and Engineering Physics, Tulane University, New Orleans, Louisiana 70118, USA

**Abstract**

We have studied the effect of spin-orbital coupling (SOC) on electronic transport properties of the thermoelectric material $\beta$-K$_2$Bi$_8$Se$_{13}$ via magnetoresistance (MR) measurements. We found that the strong SOC in this material results in weak antilocalization (WAL) effect, which can be well described by the three-dimensional weak localization model. The phase coherence length extracted from theoretical fitting exhibits a power-law temperature dependence with an exponent around 2.1, indicating that the electron phase dephasing is governed by electron – transverse phonon interactions. Like in topological insulators, the WAL effect in $\beta$-K$_2$Bi$_8$Se$_{13}$ can be quenched by magnetic impurities (Mn) but is robust against non-magnetic impurities (Te). Although our magnetotransport studies do not provide any evidences for topological surface states, our analyses suggest that SOC plays an important role in determining thermoelectric properties of $\beta$-K$_2$Bi$_8$Se$_{13}$.

Corresponding author: zmao@tulane.edu

# 1. Introduction

The topological insulators (TIs) discovered in recent years have attracted enormous attentions. They are characterized by unique gapless surface states inside the bulk energy gap, which result from the inverted band structure owing to strong spin-orbit coupling (SOC). The surface states of TIs exhibit spin helical massless Dirac Fermion structure and are protected by time-reversal symmetry. Such unique surface states result in dissipationless surface transport, which can be observed in 3D mesoscopic samples [1, 2].

The strong SOC in TIs also results in another remarkable property in transport properties, *i.e.* weak anti-localization (WAL) effect [1, 2], which comes from quantum interference of scatted electron waves. WAL is not an effect possessed only by TIs, but a generic characteristic of the materials with strong SOC. WAL has been observed in many materials, such as Mg/Ag film[3] and Ti-Al-Sn alloy[4], prior to the discovery of TIs. WAL can evolve into weak-localization (WL) when SOC is tuned weak [3, 4]. WL usually occurs in a system at low temperatures where thermal perturbations are remarkably suppressed and electrons are mostly scattered elastically with preserved phase coherence. The interference of these scattered electron waves gives rise to the quantum correction to conductance [3-5]. In the presence of time-reversal invariant perturbations, such as scattering from static non-magnetic impurities, for each possible backscattering trajectory, there exists a time-reversal symmetric path which occurs with the same probability. Their corresponding electron waves interfere constructively, which leads to enhanced probability of backscattering and thus decreases conductivity. However, for the case of WAL, the electron's spin is locked to its momentum due to strong SOC, the spin part of the electron wavefunction can no longer be ignored, and a phase difference of $2\pi$ in spin states is produced for a pair of time-reversed backscattering trajectories. Owing to the nature of the half integer electron spin, a $2\pi$ phase difference would reverse the sign of the interference term, thereby suppressing backscattering and resulting in enhanced conductance. Such quantum conductance corrections in WL and WAL can be suppressed by magnetic field. The magnetic flux through a pair of time-reversed backscattering loops produces a relative phase shift between them, thereby affecting the interference amplitude and leading to negative magnetoresistance (MR) for WL and positive MR for WAL, respectively. Specifically, MR vs. field shows a dip in low field region for the case of WAL, but a cusp for WL. [3-5] Moreover, the time-reversal-broken perturbations, such as magnetic impurities, can also effectively break the coherence of spin wavefuction, and lead to a crossover from WAL to WL in systems with strong SOC such as TIs. [6]

Three-dimensional (3D) topological insulators such as $Bi_2Te_3$ and $Bi_2Se_3$, are all actually thermoelectric materials [7, 8]. It has been proposed that the exotic surface states in TIs may be associated with their high thermoelectric efficiency [9-11], which motivates extensive search for new TI materials among thermoelectric materials. In this article, we report studies on the SOC effect of thermoelectric material $\beta$-$K_2Bi_8Se_{13}$, which has been reported to have high thermoelectric efficiency, with the figure of merit $ZT \sim 0.22$ at 300K [12]. Our focus is to clarify the role of SOC in the electronic transport mechanism of this material and show its potential relevance with current TIs. $\beta$-$K_2Bi_8Se_{13}$ crystalizes in monoclinic (space group $P2_1/m$) structure which is built up with the Bi-Se framework, as presented by a projection to $ac$ plane in Fig. 1. Such a Bi-Se framework repeats itself and extends along the crystallographic $b$-axis, forming quasi-one dimensional (1D) structure. SOC is naturally expected to be strong in this material since its structure includes heavy atoms Bi and Se. Through systematic MR measurements on undoped and doped single crystals of $\beta$-$K_2Bi_8Se_{13}$, we will demonstrate that this material indeed possesses strong SOC effect, which is manifested in the WAL behavior. Furthermore, we also observed a crossover from WAL to WL-like behavior when this material is doped by magnetic impurity Mn. In contrast, non-magnetic Te doping in this material preserves WAL and enhances SOC. These results, while they do not provide evidences for topological surface states, suggest that the SOC may play an important role in determining thermoelectric properties.

2. **Experiment**

Both undoped and Mn/Te doped $\beta$-$K_2Bi_8Se_{13}$ single crystals studied in this work were synthesized using a self-flux method. The high purity elements of K, Bi, and Se (and Mn or Te for doped samples) with stoichiometric ratio were loaded into an alumina crucible and sealed into a quartz tube under high vacuum. The mixture was then heated to 850 °C and kept at this temperature for 48 hours, followed by slowly cooling down to 200 °C. The thin, needle-like single crystals with typical diameter of ~10-20 $\mu$m could be easily obtained, as shown in the inset of Fig. 2.

The composition and structure of $\beta$-$K_2Bi_8Se_{13}$ single crystals were examined by energy-dispersive x-ray spectroscopy and x-ray diffraction analyses, respectively. The resistivity was measured using a standard four-probe method with the current flow along the needle direction (*b*-axis in Fig. 1)

3. Results

3.1 WAL in $\beta$-$K_2Bi_8Se_{13}$.

The temperature dependence of resistivity of a typical needle-like $\beta$-$K_2Bi_8Se_{13}$ single crystal is presented in Fig. 2. Data reproducibility was confirmed by measuring several different samples. As seen in Fig. 2, $\beta$-$K_2Bi_8Se_{13}$ is metallic despite its relatively large band gap ~0.59 eV [12], which may be due to non-stoichiometric self-doping effects. However, its resistivity ratio between 300K and 2K, *i.e.* $\rho(300K)/\rho(2K)$, is as small as 2.8, suggesting bad metallic behavior. This is consistent with the result reported previously by Chung *et al* [12]. As noted above, strong SOC is expected in $\beta$-$K_2Bi_8Se_{13}$ owing to the presence of heavy Bi and Se atoms. Thus WAL is anticipated to occur at low temperatures due to the quantum correction to conductance according to the introduction given above. As shown in Fig. 3(a), the MR, defined as $\Delta\rho/\rho = [\rho(B)-\rho(0)]/\rho(0)$, is positive. At low temperatures, a sharp dip in MR is seen at low field, which implies the presence of WAL effect induced by strong SOC [3, 4]. With increasing temperature *T*, the MR dip gradually broadens and finally disappears [Fig. 3(c)], leaving a quadratic field dependence at high temperatures. Such a crossover transition can be attributed to the suppression of WAL by the enhanced inelastic scatterings at high temperatures which break quantum coherence.

To make quantitative analyses of the WAL in $\beta$-K$_2$Bi$_8$Se$_{13}$, we assume that the low field MR in such a nonmagnetic system can be expressed as the classical orbital MR due to Lorentz effect plus the MR due to the WAL effect, *i.e.* $\Delta\rho = \Delta\rho_{orb} + \Delta\rho_{WAL}$. The Lorentz effect dominates the MR at high temperature, while the WAL effect governs the MR at low temperature. The contributions from Zeeman splitting and electron-electron interactions have not been taken into consideration, since their corrections to MR are usually noticeable only at high fields. In the classical diffusion regime, Kohler's rule states that for the materials with one dominant scattering rate, the classical MR at various temperatures can be represented by a universal function $\Delta\rho/\rho(0) = F(B/\rho(0))$ [13]. Specifically, the orbital MR, $\Delta\rho_{orb}$, could be scaled to $[B/\rho(0)]^2$ in the weak field limit, *i.e.* $\Delta\rho_{orb}/\rho(0) = k[B/\rho(0)]^2$ with $k$ being a constant. Therefore, at high $T$ where only orbital MR dominates, the MR data should collapse onto a single curve when being scaled to $[B/\rho(0)]^2$, which is seen in our high-$T$ ($T > 60$K) data for $\beta$-K$_2$Bi$_8$Se$_{13}$, as shown in Fig. 3(b). Consequently, at lower $T$ where WAL effect is no longer negligible, the orbital MR component could be derived according to Kohler's rule, as shown by an example of $T = 2$ K in Fig. 3(a).

For further quantitative analyses of WAL effect, we present the measured total MR data, $\Delta\rho = \Delta\rho_{orb} + \Delta\rho_{WAL}$, in the form of $\Delta\rho/\rho(0)^2$ in Fig. 3(c). While $\Delta\rho_{orb}$ at various temperatures can be obtained using Kohler's rule as stated above, the WAL effect-induced MR, $\Delta\rho_{WAL}$, can be fitted using the WAL model. Although $\beta$-K$_2$Bi$_8$Se$_{13}$ crystallizes in quasi-1D structure (Fig. 1), the diameter of needle-like crystals used in our experiment is $\sim 20$ $\mu m$, which should be far larger than typical quantum coherence length $l_\phi$ ($\sim$ a few hundred $nm$). Therefore we restrict our analyses to a 3D model. According to the 3D WAL theory, in the limit of weak magnetic field and relative large electron diffusivity $D$ (*i.e.* $D > 2\times10^{-4}$ m$^2$ s$^{-1}$) [14], the magnetic field dependence of quantum correction to resistivity in the presence of SOC is:

$$\frac{\Delta\rho_{WAL}}{\rho(0)^2} = \frac{e^2}{2\pi^2 h}\sqrt{\frac{eB}{\hbar}}[\frac{1}{2}f_3(\frac{B}{B_\phi}) - \frac{3}{2}f_3(\frac{B}{B_i + \frac{2}{3}B_s + \frac{4}{3}B_{SO}})], \quad (1)$$

where

$$f_3(y) = \sum_{n=0}^{\infty}[2(n+1+\frac{1}{y})^{1/2} - 2(n+\frac{1}{y})^{1/2} - (n+\frac{1}{2}+\frac{1}{y})^{-1/2}], \qquad (2)$$

and $B_x = \hbar/(4eD\tau_x)$ with $x$ = SO, $i$, and $s$, referring to the characteristic field for spin-orbit, inelastic, and magnetic spin-flip scattering, respectively. The inelastic and spin-flip scattering contribute to quantum decoherence, leading to the dephasing rate $\tau_\phi^{-1} = \tau_{i\phi}^{-1} + 2\tau_{s\phi}^{-1}$; the corresponding characteristic dephasing field $B_\phi = \hbar/(4eD\tau_\phi) = B_i + 2B_s$. It can be readily seen from the above equation that the sign of MR is determined by the strength of $B_{SO}$ relative to $B_i$ and $B_s$, and strong SOC leads to WAL with negative MR.

Owing to the absence of spin scattering from magnetic impurities in nonmagnetic undoped β-$K_2Bi_8Se_{13}$, only two independent parameters, $B_{SO}$ and $B_\phi$, are involved in fitting. As shown in Fig. 3(c), the 3D WAL model describes the MR data very well. The fitting parameters $B_{SO}$ and $B_\phi$ at various temperatures are presented in Fig. 4(a). The dephasing field $B_\phi$ increases nearly by three orders of magnitude from 2 K to 30 K. Such strong temperature dependence is expected since $B_\phi$ is directly related to the inelastic scattering rate, which grows with increasing temperature. On the other hand, the characteristic field for SOC, $B_{SO}$, is roughly temperature-independent, consistent with the observation in other materials with strong SOC [15]. The magnitude of $B_{SO}$, is estimated to be ~ 1.6 T, which is nearly three orders of magnitude larger than $B_\phi$ at 2 K, indicating that the spin-orbit scattering dominates the transport at low temperatures. In addition to the fitting with the large $D$ approximation, we also carried out fitting with the diffusivity $D$ being included as one of fitting parameters. The values of D obtained from the fitting are unreasonably large, which is probably caused the considerable complexity of the equation used for the fitting (see Eq. (2.1) in Ref. [14]). Nevertheless, other parameters, $B_{SO}$ and $B_\phi$, obtained from the fitting do not show any essential differences from those obtained in the fitting with the large $D$ approximation.

From the value of $B_\phi$, we could estimate the quantum coherence length, $l_\phi$, using $B_\phi = \hbar/(4eD\tau_\phi)$ and $l_\phi = (D\tau_\phi)^{1/2}$. We find that $l_\phi$ is ~ 240 nm at $T$ = 2 K, then dramatically drops to 13 nm at 30 K. $l_\phi$ is greatly less than the sample dimension (~20$\mu m$ in diameter), consistent with the 3D WAL model. $l_\phi$ characterizes the phase relaxation length of electronic state, which is reduced by phase-decoherence processes, such as inelastic scattering and spin-flip scattering. Therefore, critical information on the low-lying electronic excitations could be revealed by $l_\phi$. Theoretically, the dephasing caused by inelastic scattering leads $l_\phi$ to follow the power law temperature dependence in 3D systems [4]. Given that both the inelastic electron-electron and electron-phonon scattering contribute to the dephasing, in the absence of spin-flip scattering, $l_\phi$ ($\propto \tau_\phi^{1/2}$) can be written as [4]:

$$l_\phi^{-2}(T) = l_\phi^{-2}(T=0) + C_{ee}T^{p_{ee}} + C_{eph}T^{p_{eph}} \qquad (3)$$

where $p_{ee}$ and $p_{eph}$ are exponents corresponding to the dephasing process due to electron-electron and electron-phonon scattering, respectively. According to the 3D transport theory [4], $p_{ee}= 3/2$, while $p_{eph}$ could be 2 or 3 for inelastic scattering by transverse or longitudinal phonon, respectively.

The fitting for $l_\phi$ to Eq. (3) is presented in Fig. 4(b). The best fit is obtained by including only one power law term with exponent ~2.1. This exponent is very close to the theoretical value of 2 for electron - transverse phonon interactions, suggesting that the electron-transverse phonon interaction dominates the inelastic process; this is probably due to larger sound velocity of longitudinal phonons [16]. The electron-electron interaction is absent from our fitting ($C_{ee}$ = 0), which is reasonable since the electron-electron interaction usually dominates the transport in sub-mK temperature region in 3D systems [4]. On the contrary, in the case of 2D or 1D system, the inelastic electron-electron interaction significantly contributes to the phase relaxation [4], as seen in various thin film or nanowire samples [15-19].

### 3.2 Doping effects on WAL in $\beta$-K$_2$Bi$_8$Se$_{13}$

As observed in many SOC systems such as semiconductors composed of heavy elements [20, 21] and TIs [6], a crossover from WAL to WL-like behavior generally occurs with doping by magnetic impurities. We did similar doping studies on $\beta$-K$_2$Bi$_8$Se$_{13}$ using Mn as magnetic dopant. As expected, the transport properties are significantly altered by Mn doping. As shown in Fig. 5(a), the temperature dependence of resistivity for the 5% Mn-doped sample exhibits a remarkable upturn below $T = 110$K, probably due to random spin scattering by Mn moments. Similar to other SOC systems, magnetic doping also suppresses WAL in $\beta$-$K_2Bi_8Se_{13}$, leading to a crossover for MR from positive to negative values [Fig. 5(b) and 5(c)]. As shown in Fig. 5b, negative MR at magnetic field up to 9 T is observed in the 5% Mn-doped sample from $T = 2$ K up to 50 K. An enlarged low field MR data at $T = 2$ K is presented in Fig. 5(c), where a crossover from WAL to WL-like behavior with magnetic impurity doping can be clearly seen. The negative MR in systems doped by magnetic impurities could result from several origins. First, spin scattering by randomly distributed local moments causes enhanced resistance, which can be suppressed when these moments are aligned by magnetic field. Secondly, spin scattering also destroys the quantum coherence and thus suppresses the quantum interference, leading to the breakdown of WAL. Therefore, we are unable to fit our data to the 3D WL model (*i.e.* Eq. (1)), which takes spin scattering as small perturbations and only yields quantum corrections to resistivity induced by quantum interference. The reduction of spin scattering rate due to spin polarization is not considered in this model.

In addition to magnetic Mn doping, we also studied the Te doping effect on the magneto-transport properties of $\beta$-$K_2Bi_8Se_{13}$. Te impurities act as non-magnetic scattering centers which should not cause spin scattering. As a result, the WAL effect should be preserved. Indeed, as shown in Fig. 6(a), the resistivity for the 5% Te-doped sample shows essentially similar temperature dependence as the undoped sample does. The MR measurements also revealed robust WAL effect in this sample, as evidenced by the resistivity dip at low fields [see Fig. 6(b)]. This is in sharp contrast with the 5% Mn doping effect where the transition from WAL to WL is observed. For comparison, we have added the MR data at 2K of Te-doped sample to Fig. 5c, which shows that the Te-doped sample has the nearly same MR in the < 0.3 T field range as the undoped sample. Following the procedures for the MR analyses for undoped $\beta$-$K_2Bi_8Se_{13}$, we derived the WAL-induced MR component using Kohler's rule for the Te-doped sample, and fit the data to Eq. (1). As shown in Fig. 6b, the 3D WAL model reproduces the MR data very well from 2 K up to 30 K.

Figure 6(c) shows $B_{SO}$ and $B_\phi$ at various temperatures obtained from fittings. Similar to what has been observed in undoped $\beta$-K$_2$Bi$_8$Se$_{13}$, in the 5% Te doped sample, $B_\phi$ varies strongly with temperature, while $B_{SO}$ is roughly $T$-independent. Compared to the undoped sample, $B_{SO}$ for the 5% Te doped sample rises by 0.2 T, suggesting enhanced SOC strength, consistent with the fact that Te is heavier than Se. This result also hints that the SOC strength in $\beta$-K$_2$Bi$_8$Se$_{13}$ can be tuned by controlling the Te/Se ratio. Furthermore, we derived the quantum coherence length $l_\phi$ from $B_\phi$. $l_\phi \sim 75$ nm at 2 K, much shorter than $l_\phi \sim 240$ nm for undoped sample. This might be due to enhanced inelastic scattering because of crystal imperfections induced by chemical doping. As seen in the undoped sample, the temperature dependence of $l_\phi$ for the Te-doped sample could be fitted by Eq. (3) without including the inelastic electron-electron interaction term $C_{ee}T^{pee}$, as shown in the inset of Fig. 6c. The exponent for inelastic electron-phonon scattering term is ~2.2, very close to the fitted value ~2.1 for the undoped sample and the theoretical value 2 predicted for the electron-transverse phonon interactions in 3D systems, indicating that the dephasing mechanism remains unchanged for the non-magnetic Te doping.

4. Discussions

4.1 Can $\beta$-K$_2$Bi$_8$Se$_{13}$ be a topological insulator?

$\beta$-K$_2$Bi$_8$Se$_{13}$ shares many similarities with known TIs in terms of electronic transport and thermoelectric properties. The realization of TIs requires a gap with inverted band structure at a time-reversal invariant momentum point in Brillouin zone [22]. $\beta$-K$_2$Bi$_8$Se$_{13}$ is a narrow gap semiconductor with band gap ~0.59 eV [12] which is larger than those of TIs [23-29]. The bad metallicity seen in temperature dependence of resistivity (Fig. 2a) should be due to self-doping effect which pushes Fermi level to cross the bulk bands; this also occurs in known TIs [30]. The SOC arising from the Bi - Se framework in $\beta$-K$_2$Bi$_8$Se$_{13}$ is also strong, which has been manifested in the WAL effect probed in our MR measurements. In addition, our chemical doping studies on $\beta$-K$_2$Bi$_8$Se$_{13}$ also reveal similarities with known TIs. For TIs, the exotic spin-helical surface states are protected by time-reversal symmetry and are robust against time-reversal-invariant perturbations. Magnetic impurities, however, break the time-reversal symmetry and open an energy gap for the surface states, resulting in WL with negative MR when the gap is large [31]. Indeed, the sensitivity to magnetic doping has been experimentally observed in current TIs [6]. For $\beta$-K2Bi8Se13, as we described above, it exhibits similar response to magnetic and nonmagnetic impurities in MR as TIs do: while non-magnetic Te impurities preserve WAL, magnetic Mn impurities lead to the transition from WAL to WL [see Fig. 5(c)].

However, $\beta$-K$_2$Bi$_8$Se$_{13}$ may not be a TI in spite of these similarities, since its transport mechanism does not seem to be in line with that of TIs. As stated above, the inelastic scattering leads to quantum decoherence and suppresses quantum interference. Therefore measurements of electronic state relaxation can directly probe the scattering mechanism. TIs possess exotic transport properties due to its unique surface states; the scattering process for the surface states should only result from electron-electron interactions. This has been observed in thin film and nanowire samples of TIs, in which electron-electron scattering was found to have considerable contributions towards dephasing [16, 18, 32]. However, our above analyses for the undoped and 5% Te doped samples suggest the absence of phase relaxation from the electron-electron interaction; this seems to be against the properties of the topological surface states. Another observation that may stand against topological surface states is the absence of quantum oscillation in both the undoped and doped $\beta$-K$_2$Bi$_8$Se$_{13}$ samples [Fig. 5(b) and 6(b)]. As seen in various TIs such as Bi$_2$Te$_3$ [33], Bi$_2$Se$_3$[34, 35], and Bi$_2$Te$_2$Se [18], quantum oscillations are commonly observed and a berry phase of $\pi$ for surface states can be extracted from the Landau-level fan diagram. Various reasons may be responsible for the absence of quantum oscillation in our $\beta$-K$_2$Bi$_8$Se$_{13}$ samples. It could be simply due to the lack of topological surface states, or owing to short electron mean free path which prevents the formation of cyclotron orbits.

**4.2 The effects of SOC on thermoelectric properties**

Like many current TIs, $\beta$-K$_2$Bi$_8$Se$_{13}$ exhibits outstanding thermoelectric properties. Although our transport studies could not provide any evidences for topological surface states, the strong SOC revealed in our MR measurements may play an important role in determining thermoelectric properties of $\beta$-K$_2$Bi$_8$Se$_{13}$. The thermoelectric efficiency for a given material is characterized by a dimensionless figure of merit, $ZT = (\sigma S^2/\kappa)T$, where $\sigma$, $S$, $\kappa$ are conductivity, thermoelectric-power and thermal-conductivity respectively. Due to large $S$ and small $\kappa$, $\beta$-K$_2$Bi$_8$Se$_{13}$ exhibits relatively large $ZT$ [12]. It has been suggested that the complex unit cell with weakly bonded K$^+$ ion and mixed occupancy of K/Bi strongly scatter phonons and thereby reduce phonon conductivity of $\beta$-K$_2$Bi$_8$Se$_{13}$ [36]. However, the effect of SOC on thermoelectric properties has received little attention. As will be discussed below, large $S$ for this material could be associated with strong SOC.

The thermopower measures the voltage induced by temperature gradient. The temperature gradient creates electron and phonon flow which leads to diffusion thermopower $S_d$ and phonon-drag thermopower $S_g$ respectively. While $S_d$ is only associated with the electron diffusion, $S_g$ results from the flowing phonons which drag electrons through electron-phonon interactions. Since $\beta$-K$_2$Bi$_8$Se$_{13}$ has a very complex structure (Fig. 1), structure disorders such as interstitial sites, mixed or partial occupancies, and/or rattling atoms are very easy to be present and scatter phonons strongly. Consequently, $S_g$ is expected to be small due to reduced phonon flow. On the other hand, the electron transport is less affected in such a crystalline material. Therefore, the thermoelectric power of this material may presumably be governed by $S_d$ which is sensitive to effective charge carrier density and scattering mechanism. In general, SOC modifies the band structure and directly tunes the charge carriers available for thermal diffusion process. Besides, SOC can also cause imbalanced densities for electrons and holes, which further enhances $S_d$ via suppressing the thermoelectric voltage cancellation from opposite sign of electron and hole. Indeed, earlier band structure calculations found that the SOC could effectively alter the band structure of $\beta$-K$_2$Bi$_8$Se$_{13}$ and even close the bulk gap [36]. Moreover, as we demonstrated above, the scattering process is also remarkably affected by SOC, particularly at low temperatures. The backscattering is suppressed due to the WAL effect caused by strong SOC, which should give rise to enhanced $S_d$. More studies are called to clarify this potential mechanism.

## 5. Conclusion

In summary, we demonstrated strong SOC in the narrow gap semiconductor $\beta$-$K_2Bi_8Se_{13}$ through the WAL effect observed in MR measurements. The effect of magnetic and non-magnetic impurities on WAL in this material shows striking similarities as those seen in TIs. That is, magnetic impurities lead to the transition from WAL to WL, while the system preserves WAL in the presence of non-magnetic impurities. However, this material may not be a TI despite these similarities, since our transport mechanism analyses stand against a TI. From the fits of our MR data to the 3D WL theory, we find that the electron-transverse phonon interaction dominates the inelastic scattering process in $\beta$-$K_2Bi_8Se_{13}$. In addition, our discussions suggest that SOC may play an important role in determining thermoelectric properties of this material.

**Acknowledgements**

The work is supported by the NSF under grant DMR-1205469 and the LA-SiGMA program under award #EPS-1003897. The authors are grateful to A. Ruzsinzky and K. Yang for informative discussions.

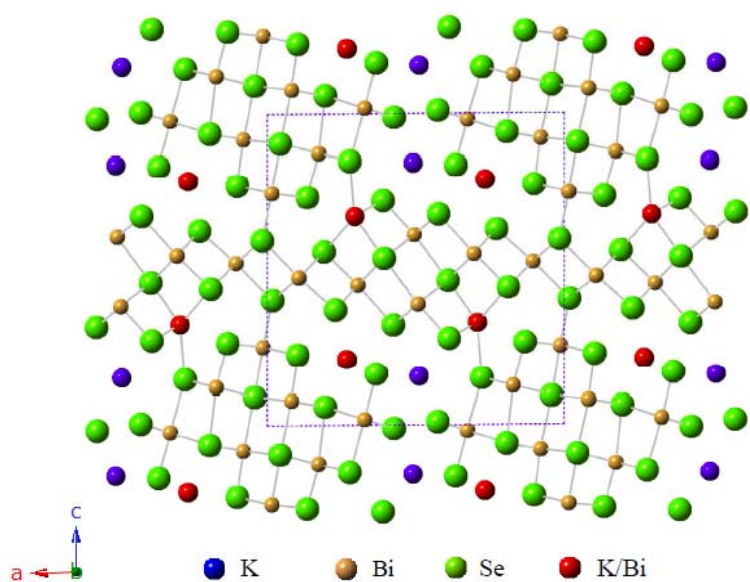

**Figure 1.** Crystal structure of $\beta$-$K_2Bi_8Se_{13}$. The Bi-Se framework extends along the *b*-axis and forms quasi-1D structure. The dashed rectangle shows the unit cell. Mixed occupancy of K/Bi atoms occurs at some sites (red sphere).

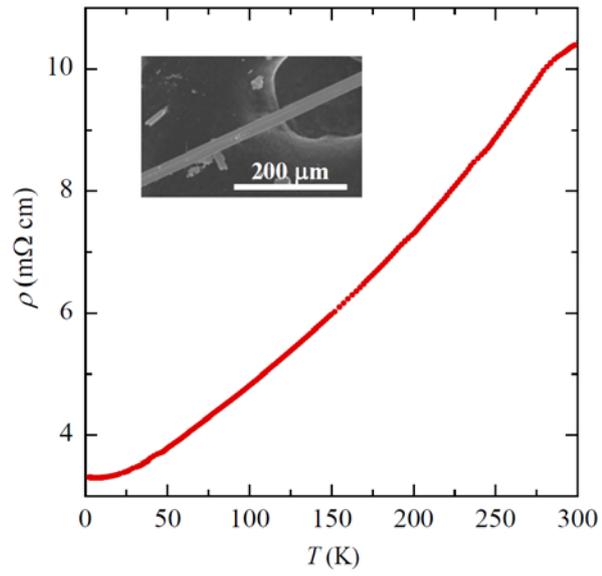

**Figure 2.** Temperature dependence of resistivity for a $\beta$-$K_2Bi_8Se_{13}$ single crystal, measured with the current flowing along the needle axis (*i.e.* the *b*-axis in Fig. 1). Inset: a scanning electron microscopy image of a typical needle-like $\beta$-$K_2Bi_8Se_{13}$ single crystal.

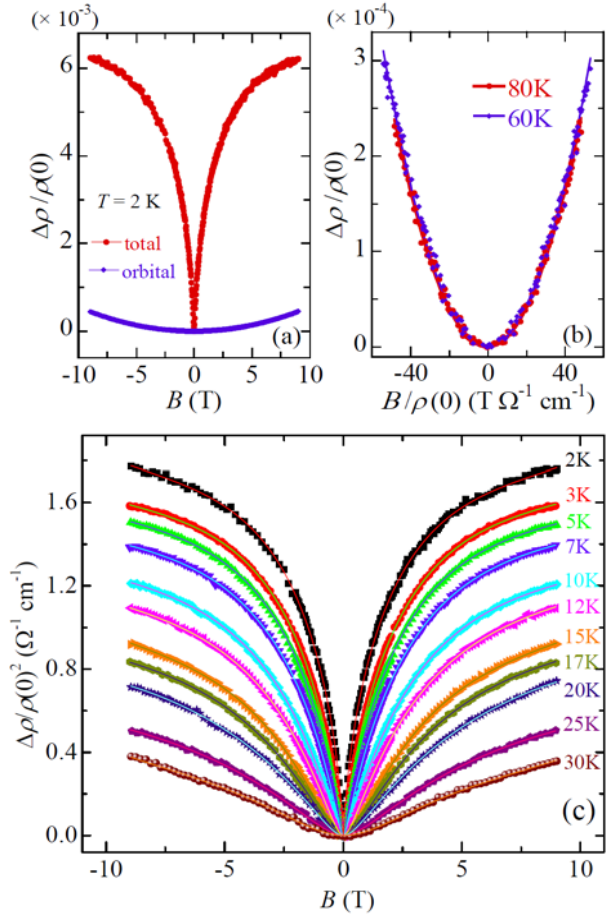

**Figure 3.** (a) MR of $\beta$-$K_2Bi_8Se_{13}$ at 2 K. The red curve shows the normalized total MR $\Delta\rho/\rho=[\rho(B)-\rho(0)]/\rho(0)$, while the blue one shows the orbital MR obtained from Kohler's rule (see text). (b) Kohler diagram showing the scaling of MR at 60 K (red) and 80 K (blue). The solid line shows the fitting of MR to $[B/\rho(0)]^2$. (c) MR of $\beta$-$K_2Bi_8Se_{13}$ at various temperatures, normalized to $\rho(0)^2$. The solid lines represent theoretical fitting based on the 3D WAL model; the fits include both components from the WAL effect ($\Delta\rho_{WAL}$) and the Lorentz effect ($\Delta\rho_{orb}$) (see text). All MR measurements were performed with magnetic field applied perpendicular to the current direction (needle direction, *b*-axis in Fig. 1).

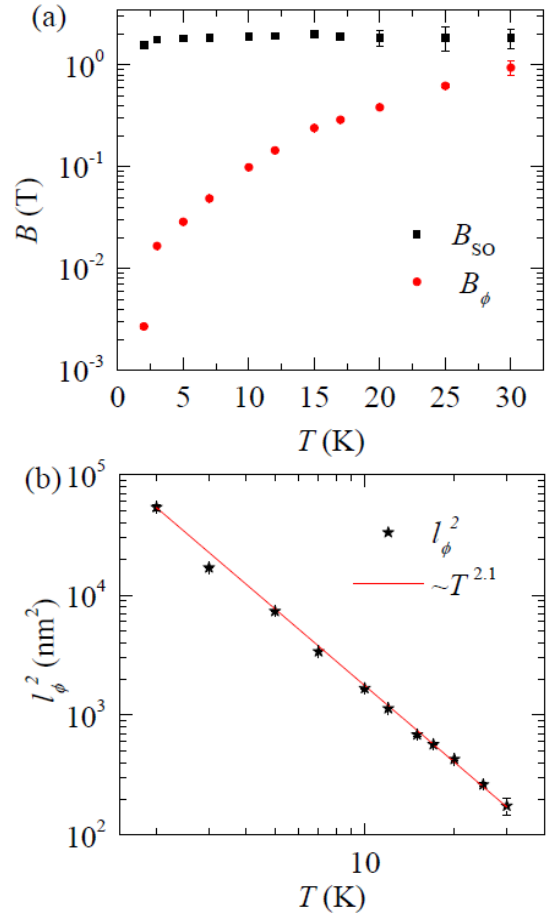

**Figure 4.** (a) Temperature dependence of characteristic fields for SOC ($B_{SO}$) and dephasing ($B_\phi$), derived from the fits of MR data to Eq. (1). (b) Temperature dependence for the square of quantum coherence length $l_\phi$, obtained from $B_\phi = \hbar/(4el_\phi^2)$. The red line shows the fit of $l_\phi^{-2}$ to $\sim T^{2.1}$ (see text).

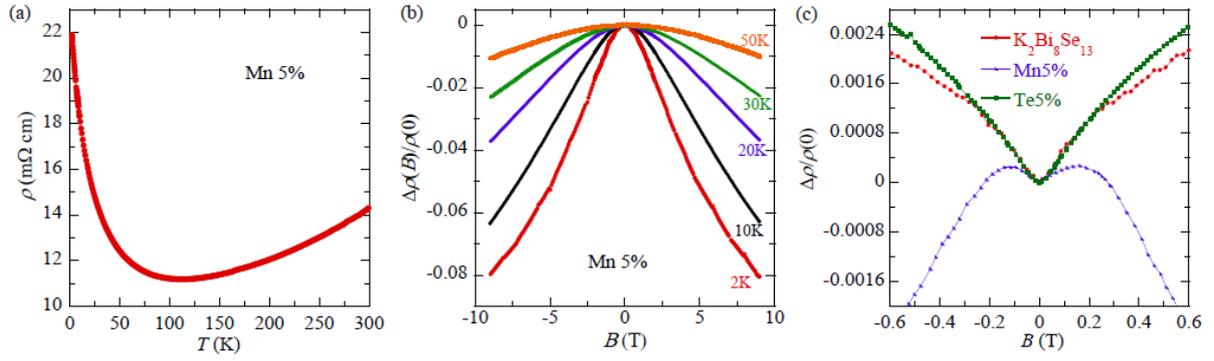

**Figure 5.** (a) Temperature dependence of resistivity for $\beta$-K$_2$(Bi$_{0.95}$Mn$_{0.05}$)$_8$Se$_{13}$. (b) Normalized MR $\Delta\rho/\rho =[\rho(B)-\rho(0)]/\rho(0)$ at various temperatures for $\beta$-K$_2$(Bi$_{0.95}$Mn$_{0.05}$)$_8$Se$_{13}$. (c) The low field MR at $T = 2$ K for undoped, Mn 5%, and Te 5% doped $\beta$-K$_2$Bi$_8$Se$_{13}$ samples.

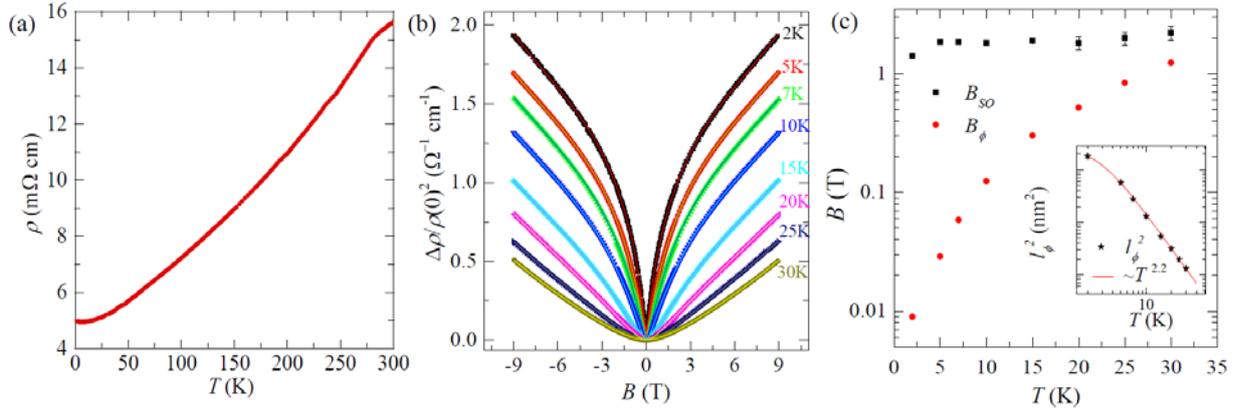

**Figure 6.** (a) Temperature dependence of resistivity for $\beta$-K$_2$Bi$_8$(Se$_{0.95}$Te$_{0.05}$)$_{13}$. (b) MR at various temperatures for $\beta$-K$_2$Bi$_8$(Se$_{0.95}$Te$_{0.05}$)$_{13}$, normalized to $\rho(0)^2$ for the purpose of fitting. (c) Temperature dependence of $B_{SO}$ and $B_\phi$ derived from the fits of the MR data to Eq. (1). Inset: Temperature dependence of the square of quantum coherence length $l_\phi$. The red line shows the fit of $l_\phi^{-2}$ to $\sim T^{2.2}$ (see text).